# Observation of Piezoelectricity in Monolayer Molybdenum Disulfide

Hanyu Zhu[1,†], Yuan Wang[1,†], Jun Xiao[1], Ming Liu[1], Shaomin Xiong[1], Zi Jing Wong[1], Ziliang Ye[1], Xiaobo Yin[1,2] and Xiang Zhang[1,2,*]

[1] NSF Nano-scale Science and Engineering Center (NSEC), 3112 Etcheverry Hall, University of California at Berkeley, Berkeley, California 94720, USA.

[2] Material Sciences Division, Lawrence Berkeley National Laboratory, 1 Cyclotron Road, Berkeley, California 94720, USA.

**Abstract:**

Piezoelectricity offers precise and robust conversion between electricity and mechanical force. Here we report the first experimental evidence of piezoelectricity in a single layer of molybdenum disulfide ($MoS_2$) crystal as a result of inversion symmetry breaking of the atomic structure, with measured piezoelectric coefficient $e_{11}$ = 2.9×10$^{-10}$ C/m. Through the angular dependence of electro-mechanical coupling, we uniquely determined the two-dimensional (2D) crystal orientation. We observed that only $MoS_2$ membranes with odd number of layers exhibited piezoelectricity, in sharp contrast to the conventional materials. The piezoelectricity discovered in single molecular membrane promises scaling down of nano-electro-mechanical systems (NEMS) to single atomic unit cell – the ultimate material limit.

Since its discovery in 1880, piezoelectricity has found a wide range of applications in actuation, sensing, and energy harvesting. Rapidly growing demands for high performance devices in micro-electro-mechanical systems (MEMS) and electronics call for nanoscale piezoelectric materials, whose potential has been recently demonstrated in mechanical resonator for low-power logic circuits, biological sensors, nano power generators, and strain-gated electronic devices (*1-10*). Reducing the size of bulk materials has been suggested to enhance piezoelectricity where quantum confinement modifies electronic states (*11, 12*). Yet when materials approach sub-nanometer scale, the large surface energy can cause piezoelectric crystalline structures to be thermodynamically unstable. For example, PbZrTiO$_3$ and BaTiO$_3$ films with few-unit-cell thickness lose spontaneous polarization (thus piezoelectricity) at room temperature (*13, 14*). This challenge has motivated theoretical investigations in novel low dimensional systems including nanotubes and single molecules (*15-17*). Layered crystals such as transition metal dichalcogenides (TMDC) have been recently theoretically proposed as promising low-dimensional piezoelectric materials (*18*), because they can retain atomic structures down to single layer (thickness less than 0.7 nm) without lattice reconstruction even in ambient conditions (*19*), and most importantly due to their broken inversion symmetry. Unique properties originated from the structural non-centrosymmetry, such as circular dichroism and second harmonic generation, have been observed in monolayer MoS$_2$ and attracted extensive research interest (*20-23*). However, experimental evidence of piezoelectricity in these 2D semiconductors has not been found so far. Here for the first time, we report converse piezoelectricity in suspended monolayer MoS$_2$ membranes. The measured piezoelectric coefficient of $e_{11}$ = 2.9×10$^{-10}$ C/m (or 0.5 C/m$^2$ when normalized by the layer thickness) is comparable to the bulk values of standard piezoelectric crystals like ZnO and AlN and an order of magnitude higher than polymeric piezoelectrics (*24*). Interestingly, we found that this molecular piezoelectricity only exists in odd number of layers in 2D crystal where the inversion symmetry breaking occurs. The angular dependence of piezoelectric response agrees with the 3-fold symmetry of the crystal and is capable of determining its atomic orientation.

The single piezoelectric coefficient $e_{11}$ can fully describe the anisotropic electromechanical coupling in monolayer $MoS_2$ due to its crystalline $D_{3h}$ symmetry (Fig. 1). The membrane has one atomic layer of molybdenum in between two identical sulfur layers packed in a hexagonal lattice. Each rhombic prismatic unit cell is asymmetrically occupied by two S atoms on the left site and one Mo atom on the right, such that an external electric field pointing from the S site to Mo site in the hexagonal lattice (armchair direction, $E_1$) can deform the unit cell by stretching the Mo-S bond and cause internal piezoelectric stress (Fig. 1B). Further analysis based on the 3-fold rotational symmetry shows the induced piezoelectric stress tensor has only one independent component $\Delta\sigma_p = \Delta\sigma_{11} = -\Delta\sigma_{22}$, which is proportional to the electric field (*25*):

$$\Delta\sigma_p = -e_{11}E_1 \tag{1}$$

To observe piezoelectricity, the conventional method measures the dipole induced by strain and has been applied to nanowires (*26*). However, the atomic thickness of single layer of $MoS_2$ significantly reduces the measurable piezoelectric charge accumulation, making the measurement very difficult. The other common technique, piezoresponse force microscopy (PFM) measures deformation with picometer sensitivity in nanometer spatial resolution (*27-29*). Unlike in conventional PFM, $MoS_2$ membrane is not coupled to the vertical electric field between tip and substrate due to its mirror symmetry along z-axis. Here, we develop a method that combines a configuration with laterally applied electric field (*30*) and nano-indentation (*31, 32*) in an atomic force microscope (AFM) to measure the piezoelectrically generated membrane stress (Fig. 1C). First, to convert in-plane stress to out-of-plane force, a suspended $MoS_2$ stripe was indented by an AFM probe and deformed. Next, piezoelectric stress was induced by an in-plane electric field applied through the lateral electrodes. This extra field-induced stress changed the load on the tip and the curvature of the cantilever, which was measured via its reflection of a laser beam. We could then quantify the piezoelectric responses of the monolayer $MoS_2$ crystal (Eq. 1).

The freestanding single layer $MoS_2$ devices were fabricated from exfoliated monolayer flakes on poly-methyl-methacrylate (PMMA). The flakes and the PMMA layer were transferred together onto hydrogen silsesquioxane (HSQ), and then monolayer $MoS_2$ in between the PMMA/HSQ layers were identified by confocal fluorescence microscopy (Fig. 2A). In order to apply electric field along armchair direction the electrodes were designed to be parallel to or at 60° with the sharply cleaved edges forming 60° or 120° corners, which was also confirmed by second harmonic generation (*22*). Suspension, mechanical clamping and electrical contact were simultaneously achieved by one-step exposure in the dash-line area through electron beam lithography (EBL), where crosslinked HSQ (amorphous $SiO_2$) composed supporting posts under $MoS_2$ membrane and PMMA was removed to open a window for Au electrodes on top (Fig. 2B). After inspecting the membrane by atomic force microscopy in non-contact mode, we measured the load and piezoelectric response as a function of depth through indentation in contact mode.

The load-indentation relation of a clamped membrane that we used to determine the material properties was deduced from finite element calculation. The width of the membrane was set much larger than its length between clamps so that the load was insensitive to variations of the shape and distance of free edges. Within the regime of small indentation, the elastic response of the membrane without electric field was described by its Young's modulus $Y^{2D}$, Poisson's ratio $v$ and pre-stress $\sigma^{2D}$ in cubic polynomial form except two geometry-dependent pre-factors (*33*). Taking $v = 0.25$ from (*18*) we found the load of the membrane could be described by (supplementary materials):

$$F = 2.1\sigma^{2D}L\left(\frac{d}{L}\right) + 3.1Y^{2D}L\left(\frac{d}{L}\right)^3 \qquad (2)$$

where $L$ is the distance between the electrodes and $d$ is the depth of indentation. Figure 3A shows the fitting of experimental data on a monolayer $MoS_2$ device with $Y^{2D} = (1.2\pm0.1)\times10^2$ N/m and $\sigma^{2D} = 45\pm5$ mN/m, agreeing well with previous *ab initio* calculation and experimental results (*18, 34, 35*). Repeatable load-indentation curves demonstrated the quality of the atomic crystalline film and the effectiveness of the clamp.

With electric field, the induced piezoelectric stress $\Delta\sigma_p$ in $MoS_2$ was treated as a perturbation during indentation, since it was two orders of magnitude smaller than the pre-stress $\sigma^{2D}$ of the suspended film according to our experimental conditions and theoretical value of electro-mechanical coupling strength. This stress provided an additional piezoelectric load $\Delta F_p$ to the film under indentation, which can be approximated as (supplementary materials):

$$\Delta F_p = 1.3(\Delta \sigma_p)d \qquad (3)$$

This additional load was measured by the tip displacement $\Delta z$ through $\Delta F_p=(k_t+k_m)\Delta z$ where $k_t$ is the stiffness of AFM cantilever, and $k_m=\partial F/\partial d$ is the stiffness of MoS$_2$ membrane derived from the load-indentation relation (Eq. 2). The numerical pre-factor 1.3 is smaller than the one preceding $\sigma^{2D}$ (2.1) in Eq. 2 as a result of the anisotropy of piezoelectric stress.

Due to the extremely small force ($\Delta F_p \sim$ pN) and displacement ($\Delta z \sim$ pm) involved, a phase lock-in scheme was employed where $\Delta \sigma_p$ was modulated by alternating electric field and $\Delta z$ was measured at the same-frequency. We kept frequency of electrical actuation ($f_p \sim$ 10 kHz) well below the fundamental resonance of the tip-film system ($f_0 \sim$ 100 kHz) so that the previous quasi-stationary mechanical model was applicable. The lock-in amplification effectively rejected non-piezoelectric higher-order effects, such as the thermal expansion by possible Joule heating, the capacitive pressure between film and substrate, and electrostriction. In order to eliminate spurious force oscillating at the same frequency of the driving voltage from static charge or contact potential difference between surfaces, we used a conductive tip and a degenerately doped silicon substrate (*29*). Both were biased with respect to membrane to balance the contact potential until the spurious force was minimized (*36*).

The piezoelectric load $\Delta F_p$ at zero depth of indentation must vanish (Eq. 3) and was used as a benchmark to verify the absence of electrostatic force (Fig 3B). At fixed driving voltage ($V$ = 1 V), fitting of the depth dependence of $\Delta F_p$ with the finite element calculation (red curve) resulted in a piezoelectric stress of $\Delta \sigma_p$ = 0.12±0.01 mN/m, to which a positive sign was assigned because the signal and the driving voltage were in phase. Meanwhile, the piezoelectric stress at fixed depth of indentation increased with ramping driving voltage (Fig. 3C), where a linear fit according to Eq. 1 gave $e_{11}$ = -2.9×10$^{-10}$ C/m, the absolute value of which was close to reported DFT calculation result (*18*).

The negative sign came from the fact that the pre-defined positive direction of electric field was anti-parallel to the x-axis in the coordinate system in Fig. 1. Therefore the anisotropy of piezoelectricity in monolayer MoS$_2$ offers a mesoscopic way to unambiguously determine its crystalline orientation, which previously can only be revealed by atomic imaging. Due to its $D_{3h}$ group symmetry, the piezoelectric coupling is a function of the crystal's azimuthal angle $\theta$ between the mirror plane containing the axis of rotation in the crystal structure and the direction of electric field (Fig. 4A):

$$\Delta \sigma_p / E = -e_{11} \cos 3\theta \tag{4}$$

Although when $\theta \neq 0$ the shear force term $\Delta \sigma_{12}$ becomes non-zero, it has no contribution to the load of tip (see supplementary materials), so we can still extract the piezoelectric stress using Eq. 3. We fabricated a series of devices based on the same single-crystal flake to study the angular dependence (Fig. 4B), with $\theta$ increased from 0° to 60°. The piezoelectric coupling of these devices (Fig. 4C) fit well to a cosine curve (red solid line) with 120° period and an angular error of less than 2°, in accordance with the crystalline orientation inferred from second harmonic generation (see supplementary materials). The change of sign from the upper devices to the lower ones allowed us to assign the atomic orientation to the 2D crystal as overlaid in the optical image.

In addition, we studied the thickness dependence of piezoelectric coefficient of 2D membranes exfoliated from natural 2H-MoS$_2$ crystals. We observed piezoelectric response occurred only for odd-layer membranes due to inversion symmetry breaking (Fig. 4D). For even-layer membranes, the contributions to piezoelectricity from alternating orientation of adjacent layers cancelled. Such results in 2D crystals marked the distinctive thickness dependence of piezoelectric coefficient from the linear scaling of conventional piezoelectric materials. The strain-gradient-induced piezoelectric coupling or "flexoelectricity" was estimated to generate ~ 0.1 pN force, more than one order of magnitude lower than the force from normal piezoelectricity (*37*), since the curvature of the indented membrane in our experiment was small. Common error sources of AFM measurement such as the hysteresis of the piezotube and the degradation of tip were constantly monitored. The angular and layer dependence of electromechanical response of the devices also provided independent evidence that indeed the signal in our measurement came from the piezoelectricity of MoS$_2$.

In conclusion, we report the first observation of molecular piezoelectricity in monolayer MoS$_2$ crystals. Unlike in bulk piezo materials, we reveal such a 2D crystal exhibits unique odd-number layer dependence due to the inversion symmetry breaking, which offers new degree of freedom to control electromechanical coupling at atomic scale. We found that angular dependence of piezoelectricity provides a mesoscopic method to probe the absolute orientation of 2D crystals which is crucial for valleytronics and edge engineering. As the flexural rigidity scales with the device thickness (*38*), 2D piezoelectric materials could greatly enhance the mechanical displacement. The robust piezoelectricity of 2D materials makes them excellent material choices for atomically thin piezoelectric devices (*39*) and enhanced electromechanical nanocomposites (*40*). With reduction in size, weight and energy consumption, we envision such 2D piezo materials will make profound impacts in ultra-sensitive sensors, nano-scale electromechnical systems, and next generation of low power electronics (*41*).

33. J. Y. Pan, P. Lin, F. Maseeh, S. D. Senturia, Verification of FEM analysis of load-deflection methods for measuring mechanical properties of thin films. *Technical Digest. IEEE Solid-State Sensor and Actuator Workshop (Cat. No.90CH2783-9)*, 70-73 (1990).

34. S. Bertolazzi, J. Brivio, A. Kis, Stretching and Breaking of Ultrathin MoS2. *Acs Nano* **5**, 9703-9709 (2011).

35. R. C. Cooper *et al.*, Nonlinear elastic behavior of two-dimensional molybdenum disulfide. *Physical Review B* **87**, (2013).

36. J. M. R. Weaver, D. W. Abraham, HIGH-RESOLUTION ATOMIC FORCE MICROSCOPY POTENTIOMETRY. *Journal of Vacuum Science & Technology B* **9**, 1559-1561 (1991).

37. A. G. Petrov, Flexoelectricity of model and living membranes. *Biochimica Et Biophysica Acta-Biomembranes* **1561**, 1-25 (2002).

38. J. A. Rogers, M. G. Lagally, R. G. Nuzzo, Synthesis, assembly and applications of semiconductor nanomembranes. *Nature* **477**, 45-53 (2011).

39. M. Lopez-Suarez, M. Pruneda, G. Abadal, R. Rurali, Piezoelectric monolayers as nonlinear energy harvesters. *Nanotechnology* **25**, 175401 (175405 pp.)-175401 (175405 pp.) (2014).

40. S. H. Zhang, N. Y. Zhang, C. Huang, K. L. Ren, Q. M. Zhang, Microstructure and electromechanical properties of carbon nanotube/poly(vinylidene fluoride-trifluoroethylene-chlorofluoroethylene) composites. *Advanced Materials* **17**, 1897-1901 (2005).

41. J. L. Arlett, E. B. Myers, M. L. Roukes, Comparative advantages of mechanical biosensors. *Nature Nanotechnology* **6**, 203-215 (2011).
12

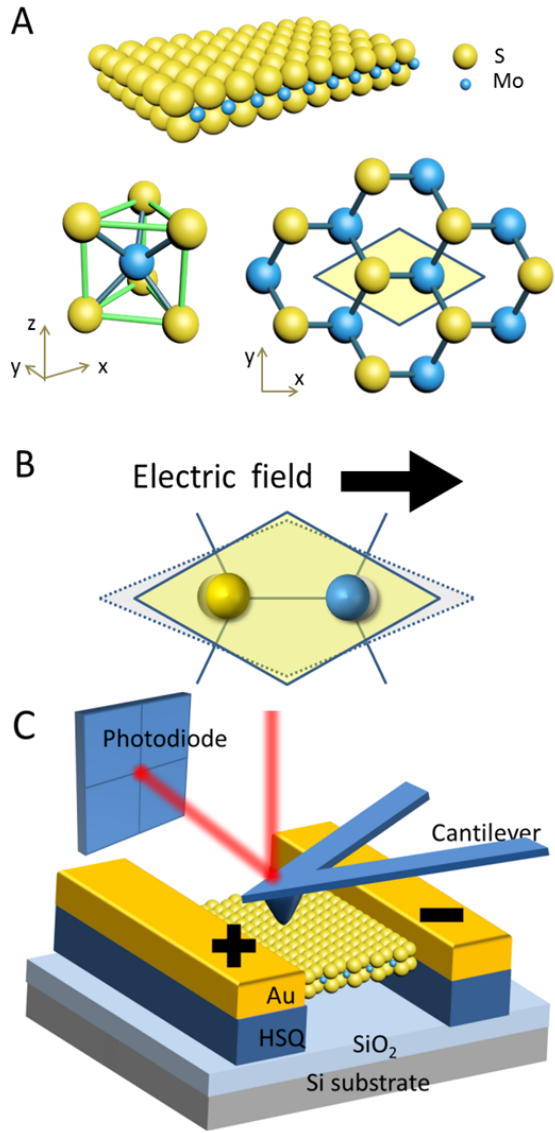

**Fig. 1**. **Probing piezoelectric property of freestanding monolayer MoS$_2$.** (**A**) A single layer of MoS$_2$ consists of S-Mo-S stacking with a total thickness of 0.6 nm. From the top view each unit cell consists of two sulfur atoms occupying the same site in the hexagonal lattice while the molybdenum atom residing in the opposite site, and therefore breaks the inversion symmetry in XY plane but preserves mirror symmetry in Z direction. (**B**) With an external electric field pointing from S-site to Mo-site, the Mo-2S dipole is stretched and the unit cell is elongated, creating compressive stress in X-direction and tensile stress

in Y-direction. **(C)** To measure the in-plane piezoelectric stress, the MoS$_2$ film was suspended on two HSQ posts, clamped underneath two Au electrodes. The film was indented by a scanning AFM probe. The induced stress changed the load on the cantilever, which was observed via the deflection of a laser beam.

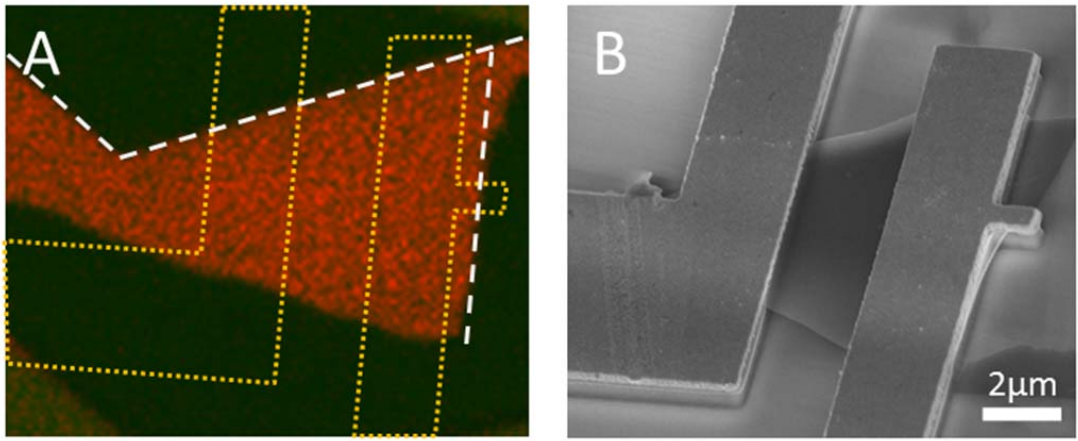

**Fig. 2. Design and characterization of the piezoelectric monolayer MoS$_2$ device. (A)** To maximize piezoelectric coupling, electrodes (outlined with yellow dot lines) were oriented parallel to the zigzag edges (white dash lines, verified by second harmonic generation measurement) of exfoliated monolayer flakes identified by confocal fluorescence microscopy. **(B)** The device was inspected by SEM to confirm that freestanding MoS2 monolayers were clamped between Au electrodes and HSQ posts, patterned through one-step electron beam lithography. The clearance between film and substrate was large (>450nm) enough to reduce the van der Waals force from the substrate during indentation.

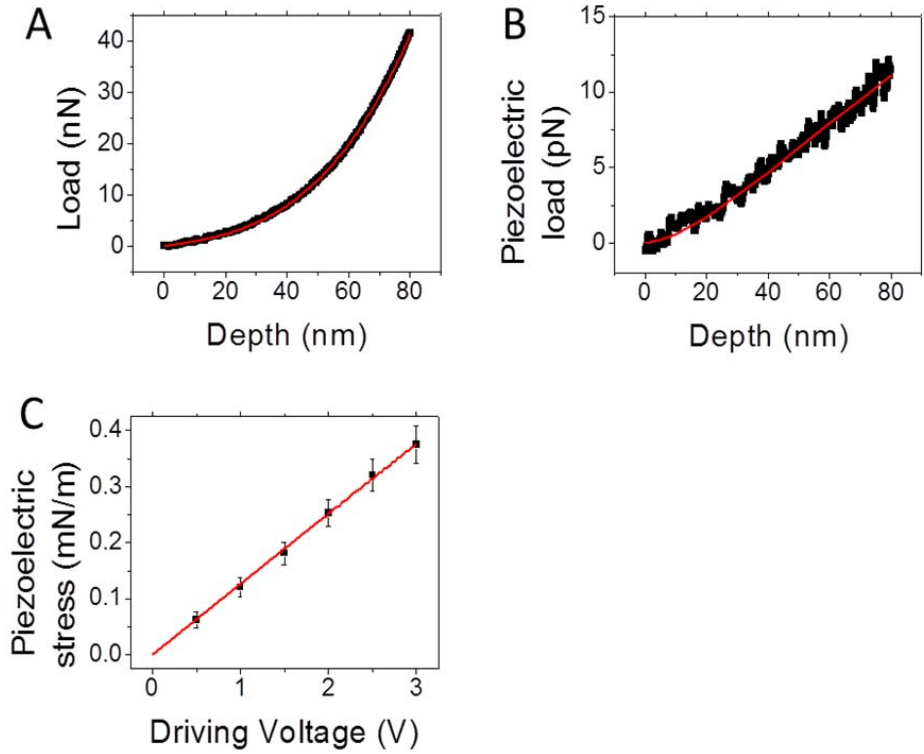

**Fig 3. Measuring piezoelectric coefficient through nano-indentation and electromechanical actuation.** (A) The mechanical properties of MoS$_2$ membranes without electric field are characterized with the load-indentation dependence. A vertical load exhibited cubic dependence on indentation depth described by linear elastic membrane theory (Eq. 2). Young's modulus ($Y^{2D}$ = 1.2×10$^2$ N/m) of monolayer MoS$_2$ retrieved from polynomial fitting was in good agreement with previous studies. The load curve also provided effective spring constant of the film as a function of depth. (B) Piezoelectrically induced load ($\Delta F_p$) on the tip at fixed driving voltage ($V$ = 1 V) was measured as a function of the depth of indentation, and the slope was the scalar piezoelectric stress ($\Delta \sigma_p$ = 0.12 mN/m) according to Eq. 3. $\Delta F_p$ was derived from the stiffness of tip/film assembly times the change of tip deflection $\Delta z$ directly measured by lock-in amplifier. **(C)** Piezoelectric stress measured at various driving voltage at fixed depth of indentation showed a linear dependence. The piezoelectric coefficient of monolayer $e_{11} = 2.9\times10^{-10}$ C/m was calculated from the slope.

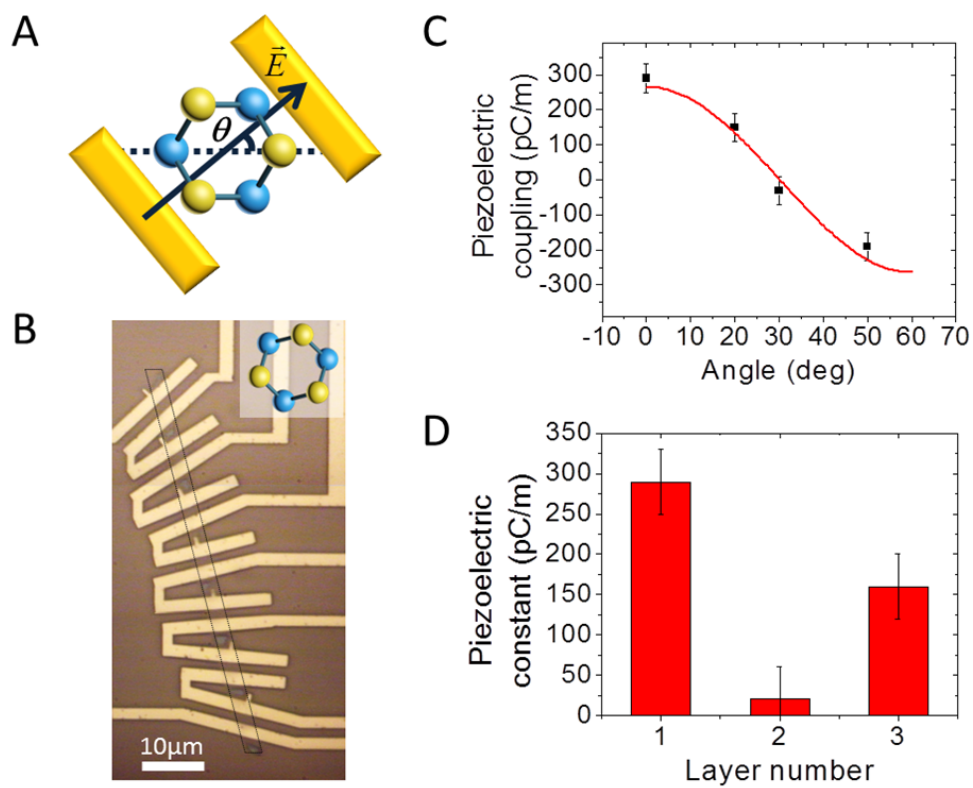

**Fig 4. Angular dependence of piezoelectric response in monolayer MoS$_2$.** **(A)** The rotation of crystal with respect to the electric field and mechanical boundary was achieved by patterning the electrodes at different angles. **(B)** Optical image of multiple electrode pairs integrated on a long stripe of MoS$_2$ rotated by 10° at each section. The relative rotation between the first and last device was 60°, so that the piezoelectric effect should reverse the sign as the alignment of electric field to Mo-2S dipole changes from parallel to anti-parallel. **(C)** Measured piezoelectric coupling strength followed the cos(*3θ*) dependence as predicted from the crystalline 3-fold symmetry. The sign change was observed from the phase read-out of piezoelectric signal through lock-in amplifier. This distinguished our low-frequency electric method from second harmonic generation which typically only gives the amplitude information, such that the crystalline orientation as shown in the inset of **(B)** could be uniquely determined without resorting to atomic

images. **(D)** Measured piezoelectric coefficient in 1-, 2- and 3-layer $MoS_2$ membranes showed that only odd-layer exhibited significant coupling strength due to their broken inversion symmetry.